\documentclass[12pt]{article}

\usepackage{amssymb}
\usepackage{epsfig,color}
\usepackage{pstricks,graphicx,epsfig,color,amssymb,amsmath,amscd}
\usepackage{cite}
\usepackage[]{graphicx}

\usepackage{makeidx}
\newcommand{\bi}{\bibitem}
\newcommand{\be}{\begin{eqnarray}}
\newcommand{\ee}{\end{eqnarray}}
\newcommand{\rar}{\rightarrow}

\usepackage[]{caption}
\def\mc{\mathcal}
\def\be{\begin{equation}}
\def\ee{\end{equation}}
\def\mpl{m_{Pl}}
\def\mn{{\mu\nu}}
\def\D{\mc D}
\def\-g{\sqrt{-g}}
\renewcommand\rho{\varrho}

\begin{document}


\title{Cosmological evolution in $R^2$ gravity}
\author{E.V. Arbuzova$^a$, A.D. Dolgov$^{b,c,d}$, L. Reverberi$^{b,c}$}

\maketitle
\begin{center}
$^{a}$Department of Higher Mathematics, University ”Dubna”\\
Universitetskaya ulitsa, 19, 141980 Dubna, Russia\\
$^{b}$Dipartimento di Fisica, Universit\`a degli Studi di Ferrara\\
Polo Scientifico e Tecnologico - Edificio C, Via Saragat 1, 44122 Ferrara, Italy\\
$^{c}$Istituto Nazionale di Fisica Nucleare, Sezione di Ferrara\\
Polo Scientifico e Tecnologico - Edificio C, Via Saragat 1, 44122 Ferrara, Italy \\
$^{d}$Institute of Theoretical and Experimental Physics\\
Bolshaya Cheremushkinskaya ul. 25, 113259 Moscow, Russia.
\end{center}



\begin{center}
\centering
\begin{table}
\begin{tabular}{r l}
 \textbf{e-mail:} & arbuzova@uni-dubna.ru\\
& dolgov@fe.infn.it\\
& reverberi@fe.infn.it
\end{tabular}
\end{table}
\end{center}

\begin{abstract}
The universe evolution during the radiation-dominated epoch in the $R^2$--extended gravity theory
is considered. The equations of motion for $R$ and $H$ are solved analytically and numerically. The
particle production rate by the oscillating curvature is calculated in one-loop approximation and the
back-reaction of particle production on the evolution of $R$ is taken into account. Possible implications
of the model for cosmological creation of non-thermal dark matter is discussed.
\end{abstract}

\newpage


\section{Introduction}
During the last two decades the accumulated astronomical data have unambiguously proved that the universe expansion
is accelerated. The data include the observation of the large scale structure of the universe,
the measurements of the angular fluctuations of the cosmic microwave background radiation, the determination of the universe age
(for a review see~\cite{cosm-prmtr}), and especially the discovery of the dimming of distant Supernovae~\cite{Nobel_2011}.

The driving force behind this accelerated expansion in unknown. Among possible explanations, the most popular is probably
the assumption of a new (unknown) form of cosmological energy density with large negative pressure, $ P < -\rho/3$, the so-called dark energy, for a review see e.g.~\cite{DE_Peebles_Ratra}. 

A competing mechanism to describe the accelerated expansion is represented by gravity modifications at small curvature, the so-called $f(R)$-gravity theories, as suggested in ref.~\cite{grav-mdf}. In these theories the standard
Einstein-Hilbert Lagrangian density, proportional to the scalar curvature $R$, is replaced by a function
$f(R)$, so the usual action of General Relativity acquires an additional term:
\be
S = -\frac{m_{Pl}^2}{16\pi} \int d^4 x \sqrt{-g}\,f(R)+S_m=
-\frac{m_{Pl}^2}{16\pi} \int d^4 x \sqrt{-g}\,\left[R+F(R)\right]+S_m\, ,
\label{A1}
\ee
where $m_{Pl}= 1. 2 2\cdot 10^{19}$ GeV is the Planck mass and $S_m$ is the matter action.

The original version of such models suffers from a strong instability in presence of gravitating 
bodies~\cite{DolgKaw} and because of that more complicated functions $ F(R) $ have been 
proposed~\cite{Starob, HuSaw, ApplBatt,Noj-Odin-2007}, which are free from the mentioned exponential 
instability. It was claimed in an earlier paper~\cite{odin-instab} that the instability could be eliminated by
a mere addition of terms of the type $R^q/m_q^{2(q-1)}$ to the action, where $m_q$ is a constant parameter 
with dimension of mass. To some extent such terms may suppress the instability for systems with
relatively high mass density, $\rho > 1 {\rm g/cm^{3}}$, but for less dense systems e.g. for those with
$\rho \sim 10^{-24} {\rm g/cm{^3}}$ this mechanism demands too large coefficients in front of $R^q$ (small $m_q$),
which are at odds with big bang nucleosynthesis.

%

Though free of instability~\cite{DolgKaw}, the models proposed in~\cite{Starob, HuSaw, ApplBatt} possess
another troublesome feature, namely in a cosmological situation they should evolve from a singular 
state in the past~\cite{appl-bat-08}. Moreover, it was found in refs.~\cite{frolov, Arb_Dolgov}
that in presence of matter, a singularity may arise in the future if the matter density rises with time; such future singularity is unavoidable, regardless of the initial conditions, and is reached in a time which is much shorter than the cosmological one.
In the standard Friedmann cosmology, in which the energy density decreases with time, a future singularity may appear
in several scenarios (phantom cosmology, quintessence models, \dots), but not in models~\cite{Starob, HuSaw,ApplBatt}, as shown in ref.~\cite{Arb_Dolgov}.
This statement is in straight disagreement with earlier papers~\cite{odin-future-sing}.

The subject of modified gravity includes about $10^3$ papers at the present time and it is impossible to
quote many relevant works. For a review one may look into references~\cite{App-Bat-Star,NojOd,clifton}.

The aforementioned problems can be cured by adding to the action an $R^2$-term,
which prevents from the singular behavior both in the past and in the future.
In the present work we study the cosmological evolution of the Universe in a theory with
only an additional $R^2$ term in the action, neglecting other terms which have been introduced
to generate the accelerated expansion in the contemporary universe. The
impact of such terms is negligible in the limit of sufficiently large curvature, $|R|\gg |R_0|$, where
$R_0$ is the cosmological curvature at the present time. 

In other words, we study here the cosmological evolution
of the early and not so early universe in the model with the action:
\be
S = -\frac{m_{Pl}^2}{16\pi} \int d^4 x \sqrt{-g} \left(R-\frac{R^2}{6m^2}\right)+S_m\,,
\label{A-R2}
\ee
where $m$ is a constant parameter with dimension of mass.\\
Cosmological models with an action quadratic in the curvature tensors were pioneered
in ref.~\cite{Gur-Star}. Such higher-order terms appear
as a result of radiative corrections to the usual Einstein-Hilbert action after taking
the expectation value of the energy-momentum tensor of matter in a curved background.
In such models, like for instance the Starobinsky model~\cite{Starobinsky_1980}, the universe may have experienced an exponential (inflationary)
expansion without invoking phase transitions in the very early universe. This model has a graceful exit to matter-dominated stage which is induced by the new scalar degree of freedom, the \textit{scalaron} (curvature scalar), which becomes a
dynamical field in $R^2$-theory.

The reheating process, due to gravitational particle production from scalaron (curvature scalar)
oscillations, leads to a transition to a Friedmann-like universe. These features of the model are thoroughly discussed, for instance,
in refs.~\cite{Vilenkin_1985,Mijic_Morris_Suen_1986,Suen_Anderson}. 
Cosmological dynamics of fourth-order gravity were investigated in several works, see e.g.\cite{Carloni_Dunsby_Troisi_2009} and references therein.

A somewhat similar study was performed in ref.~\cite{davidson} where a version of massive Brans-Dicke (BD)
theory without kinetic term (i.e. with BD parameter $\omega =0$) was considered. The Hubble parameter and
curvature demonstrate oscillating behavior which resembles the one found in our paper (and earlier in many others),
but quantitative features are very much different.


Beside $R^2$-terms, terms containing the Ricci tensor squared $ R_{\mu\nu} R^{\mu\nu}$ are induced by radiative corrections as well, and with similar magnitude. However, the natural magnitude of such
radiatively induced terms is quite small. The characteristic mass parameter, in fact, is of the order of the Planck mass
in both cases, which makes this situation non-interesting for applications discussed below. On the other hand, $R^2$ cosmology
(without  $ R_{\mu\nu} R^{\mu\nu} $) has been considered in the literature with much larger magnitude of $R^2$ than the natural
value from radiative corrections. The assumption of large $R^2$ terms is made \textit{ad hoc} to formulate a model which could,
for instance, cure singularities. We follow the spirit of those works. However, it could be worthwhile to study the
consequences of more complicate models with both $R^2$ and $ R_{\mu\nu} R^{\mu\nu} $ terms. It may be a subject for
future investigation.

The paper is organized as follows. In sec.~\ref{sec:eqs_of_motion} the cosmological equations modified by the presence of
$R^2$-term are presented and solved both analytically and numerically in the case of a radiation-dominated (RD) universe.

We study two equivalent sets of equations,
for the time-time component of the (modified) Einstein equations and for their trace. In sec.~\ref{s-part-prod} particle
production by the external oscillating gravitational field is studied. First, we derive the equation of motion for the
evolution of $R$ with the account of the back-reaction from particle production. This leads to an exponential
damping of the oscillating part of $R$, while the non-oscillating "Friedmann" part remains practically undisturbed.
The particle production influx into the cosmological plasma is estimated in the case of a massless, minimally-coupled scalar field.
In the conclusions we discuss possible implications of the scenario, in particular for heavy
supersymmetric dark matter.

\section{Friedmann-like Universe in $R^2$ Gravity}\label{sec:eqs_of_motion}

We consider the theory with action (\ref{A-R2}) and use the conventions
$\eta_\mn = \text{diag}(+,-,-,-)$,
$R^\alpha_{\,\,\mu\beta\nu}=\partial_\beta\Gamma^\alpha_{\mn}+\cdots$ and $R_\mn = R^\alpha_{\,\,\mu\alpha\nu}$.
We also use the natural units $\hbar=c=k_B=1$ and define the Planck mass as $m_{Pl}^2 = G_N^{-1}$.

The modified Einstein equations for this theory read
\be
 R_{\mn} - \frac{1}{2}g_{\mn} R -
 \frac{1}{3m^2}\left(R_{\mn}-\frac{1}{4}R g_{\mn}+g_{\mn}\mc D^2-\mc D_\mu\mc D_\nu\right)R
 =\frac{8\pi}{\mpl^2}T_\mn\,, \label{field_eqs}
\ee
where $\D^2\equiv g^\mn \D_\mu\D_\nu$ is the covariant D'Alembert operator. We assume the
Friedmann-Robertson-Walker metric with the interval given by
\be
ds^2 = dt^2 - a^2(t)\left[\frac{dr^2}{1-kr^2}+r^2d\vartheta^2+r^2\sin^2\vartheta\,d\varphi^2\right]\,.
\label{FRW}
\ee
In what follows we will neglect the three-dimensional space curvature\footnote{This is a very good approximation, at least during the
radiation-dominated epoch.}, hence setting $k=0$.\\
The curvature scalar $R$ is expressed through the Hubble parameter $H = \dot a/a$ as
\be
R=-6\dot H-12H^2\,.
\label{R-of-H}
\ee
Therefore, the time-time component of eq.~(\ref{field_eqs}) reads
\be
\ddot H+3H\dot H - \frac{\dot H^2}{2H}+\frac{m^2 H}{2} = \frac{4\pi m^2}{3\mpl^2 H}\rho\,,
\label{timetime}
\ee
where over-dots denote derivative with respect to physical time $t$.

Taking the trace of eq.~(\ref{field_eqs}) yields
\be
\ddot R + 3H\dot R+m^2\left(R+\frac{8\pi}{\mpl^2}T^\mu_\mu\right)=0\,.
\label{trace}
\ee
This equation is a sort of Klein-Gordon equation for a homogeneous 
scalar field (the ``scalaron'') of
mass $m$, with a source term proportional to the trace of the energy-momentum
tensor of matter. The General Relativity (GR) case 
may be recovered when $m\rar \infty$.
In this limit we expect to obtain the usual algebraic relation between the curvature scalar and the trace of the energy-momentum
tensor of matter:
\be
m_{Pl}^2 R_{GR} = - 8\pi T_\mu^\mu\,.
\label{G-limit}
\ee
However, unlike the usual GR, in higher-order theories curvature and matter are related to each other through a differential equation, not simply algebraically. Therefore, the theory may approach GR as $m\rightarrow \infty$ in a  non-trivial way or even not approach it at all.

For a perfect fluid with relativistic equation of state $P~=~\rho/3$,
the trace of the energy-momentum tensor of matter $T^\mu_\mu$
vanishes and $R$ satisfies the homogeneous equation. The GR solution $R=0$ satisfies this equation, but if one assumes that
neither $R$ nor $\dot R$ vanish initially, the general solution for
$R$ will be an oscillating function with a decreasing amplitude. The decrease of the amplitude is induced by
the cosmological expansion (the second term in eq. (\ref{trace})) and by particle production by the
oscillating gravitational field $R(t)$. The latter is not included in this equation and
will be taken into account below in sec.~\ref{s-part-prod}.

It can be easily shown that the left-hand side of  eq.~(\ref{field_eqs}) is covariantly conserved,
which in turn implies the covariant conservation
of the energy-momentum tensor of matter. The latter allows to write the evolution equation for the matter content, assuming it
to be a perfect fluid with energy density $\rho$ and pressure $P$:
\be
\dot\rho+3H(\rho+P)=0\,.
\label{energy_density_evolution}
\ee
As is well known, only two of equations (\ref{timetime}), (\ref{trace}), and (\ref{energy_density_evolution})
are independent.

From eq.~(\ref{energy_density_evolution}) it follows that relativistic matter, having equation of state $P=\rho/3$, satisfies
\begin{equation}\label{rho_evol}
\dot \rho_\text R + 4H\rho_\text R=0\,.
\end{equation}

In what follows  we will use either the set of eqs. (\ref{timetime}) and (\ref{rho_evol}) or
the set (\ref{R-of-H}) and (\ref{trace}) as the basic equations. They are of course equivalent
but their numerical treatment may be somewhat different.

There is a possibility of gravitational particle production, which may non-trivially affect the
solutions of the above equations. In first approximation, however, we neglect such contributions,
which will be dealt with later on in the final part of this paper.

\subsection{Evolution equations in dimensionless form}

During most of its history (in terms of redshift, not time) the universe was dominated by relativistic matter, or in other words, 
it was radiation-dominated (RD). An exception to the RD regime is of course the inflationary stage with vacuum-like energy density and the 
universe heating epoch at the end of inflation when most probably the (non-relativistic) matter dominated (MD) regime was realized. 
After that, relativistic matter was dominant till redshifts of order of $10^4$, which corresponds to the moment of the matter-radiation equality.
From the observations of the abundances of light elements we know for sure that at the time of big bang nucleosynthesis
(BBN) the universe was quite precisely dominated by relativistic matter.
If earlier in the course of the universe expansion and cooling down there were first order phase transitions in the primeval
plasma, relatively short periods of vacuum-like matter dominance could have taken place. We also expect some (small)
corrections to the RD regime due to the existence of particles in the plasma with masses comparable to the universe temperature,
and because of the trace anomaly in the energy-momentum tensor of matter which leads
to $T_\mu^\mu \neq 0$ even for massless particles.
The universe might have even been in a practically pure MD regime after the post-inflationary RD stage. This regime
could have been created by primordial black holes which evaporated early enough to bring the universe back to
the normal RD epoch~\cite{pbh-md}. The last possibility is especially interesting in the case of $R^2$-inflation since
the initial oscillations of $R$ would be damped due to particle production at the end of inflation, and the GR solution
could be restored.
All these deviations from the pure RD regime would in turn induce deviations from the GR solution ($R=0$) and
give rise to oscillations of R even if they were initially absent.

In what follows, we study the cosmological evolution in the $R^2$-theory assuming rather
general initial conditions for $R$ and $H$ and dominance of relativistic matter. The MD regime will be studied elsewhere.

It is convenient to rewrite the equations in terms of the dimensionless quantities \mbox{$\tau=H_0\,t$}, $h=H/H_0$, $r=R/H_0^2$,
$y=8\pi\rho/(3\mpl^2 H_0^2)$, and $\omega=m/H_0$,where $H_0$ is the value of the Hubble parameter at some initial time $t_0$. Thus the following
two equivalent systems of equations are obtained:
\begin{equation}\label{sys:hubble_evolution}
\begin{cases}
h'' + 3h h' - \cfrac{h'^2}{2h}+\cfrac{\omega^2}{2}\cfrac{h^2-y}{h}=0\,,\\
y' + 4hy = 0\,,
\end{cases}
\end{equation}
and
\begin{equation}\label{sys:curvature}
\begin{cases}
r''+3hr'+\omega^2r=0\,,\\
r+6h'+12h^2=0\,.
\end{cases}
\end{equation}
Here  prime indicates derivative with respect to dimensionless time $\tau$. If we impose the ``natural'' relativistic initial conditions:
\begin{equation}\label{eq:initial_conditions_1}
 \begin{aligned}
  &\tau_0 = 1/2\,,\\
&h_0=1\,,\\
&h'_0=-2\,,\\
&y_0=1\,,
 \end{aligned}
\end{equation}
we find that there exists the exact solution $h=1/(2\tau)$, $y=1/(4\tau^2)$, and $r=0$,
which corresponds to the usual general relativity, but this solution may deviate from GR because of deviations of the real 
expansion regime from the purely relativistic one. In principle, depending on the initial conditions, the
solutions may oscillate around some value of $h\tau$, which may itself be different
from $1/2$, and in fact this is what we will find later on.

Any non-negligible deviation from the GR solution may lead to observable effects, and hence to
observational constraints on $m$, the only free parameter of the model.
We will tackle these problems below both analytically and numerically.

\subsubsection{Approximate Analytical Solutions}\label{sec:analytical_estim_no_back_react}

First we assume that the deviations from GR are small and expand
$h=1/(2\tau) +h_1$ and $y=1/(4\tau^2)+y_1$, assuming that $h_1/h \ll 1$ and $y_1/y \ll 1$, and linearize the system of equations.
It is convenient to introduce a new unknown function $z_1\equiv h_1'$, so we obtain three
first-order linear differential equations with time-dependent coefficients:
\begin{equation}\label{sys:linear_perturb}
\begin{cases}
  z_1' = -\cfrac{5}{2\tau}\,z_1+\left(\cfrac{1}{\tau^2}-\omega^2\right)h_1+\tau\omega^2 y_1\\
 h_1' = z_1\\
 y_1' = -\cfrac{1}{\tau^2}\,h_1-\cfrac{2}{\tau}\,y_1
 \end{cases}
\end{equation}
We can find an approximate analytical solution of this system in the limit of large times, or
$\omega\tau\gg 1$. In this limit we can treat the coefficients as approximately constant and find the eigenvalues
and eigenfunctions of the system of differential equations. This method essentially consists in separating "fast" and "slow" variables.

The characteristic polynomial is
\begin{equation}\label{polyn_1}
 P(\lambda)=\lambda^3+\frac{9\lambda^2}{2\tau}+ \lambda\left(\frac{4}{\tau^2}+\omega^2\right)
+ \frac{3\omega^2}{\tau} - \frac{2}{\tau^3}
\end{equation}
and the eigenvalues (for large $\omega \tau$) are approximately
\begin{equation}\label{lambda123}
 \lambda_1 \approx -\frac{3}{\tau}\,, \qquad\qquad \lambda_{2,3} \approx -\frac{3}{4\tau}\pm i\omega\,.
\end{equation}
The general solutions of the system (\ref{sys:linear_perturb}) is a linear combination of eigenvectors $V_j$:
\be
[h_1, z_1, y_1] = \sum C_j V_j \exp \left[ \int^\tau d\tau'\,\lambda_j (\tau') \right]\,,
\label{h1-z1-y1}
\ee
where $V_1 = [1,\, -3/\tau,\,1/\tau]$, $V_{2,3} = [1,\, -3/( 4\tau) \pm i\omega ,\,-1/(5\tau/4 \pm i\omega \tau^2)]$ and
\be
\exp \left[ \int^\tau d\tau'\,\lambda_1 (\tau') \right] \sim 1/\tau^3\quad{\rm and}\quad
\exp \left[ \int^\tau d\tau'\,\lambda_{2,3} (\tau') \right] \sim  e^{\pm i\omega \tau} /\tau^{3/4}\,.
\label{exp-lambda}
\ee
Since the solution must be real, one should take its real and imaginary part.

Let us note that eigenvectors $V_j$ are normalized to almost constant values, up to terms
of order $1/\tau^2$. In principle coefficients $C_j$ depend upon time but this dependence
is quite weak, $C_j \sim C_{j0} + C_{j1}/\tau^2$ and asymptotically negligible. 

The correction to the GR solution corresponding to the first eigenvalue $\lambda_1$ quickly decreases, since
$h_1^{(1)} \sim 1/\tau^3$ and $y_1^{(1)} \sim 1/\tau^4$, and so it can be asymptotically neglected.
The solutions for $h_1$ corresponding to $\lambda_{2,3}$ oscillate and decrease more slowly than the GR solution, in fact
\begin{equation}
 h_1^{(2,3)} \sim \frac{\sin(\omega\tau+\varphi)}{\tau^{3/4}}\,, \label{eq:analytical_solution}
\end{equation}
while the solution for the energy density also oscillates but drops down faster than the GR one, $y \sim \tau^{-11/4}$.
The complete asymptotic solution for $h$ has the form:
\begin{equation}\label{sol_h_linear}
h(\tau)\simeq \frac{1}{2\tau}+\frac{c_1\sin(\omega\tau+\varphi)}{\tau^{3/4}}\,.
\end{equation}
For sufficiently large $\tau$ the second term would start to dominate and the linear approximation would no longer hold. 
Below we will obtain approximate analytical solutions even in the non-linear regime, in the high-frequency limit.

However prior to that it would be instructive to find the solution for the equivalent set of
equations (\ref{R-of-H}) and (\ref{trace}) in the same approximation of small deviation from GR.
Defining as above,  dimensionless time, $\tau= H_0 t$, dimensionless Hubble parameter,
$h = H/H_0$, and dimensionless curvature $r\equiv R/H_0^2$, we rewrite these 
equations in the form:
\begin{equation}\label{trace_dimensionless}
 \begin{cases}
 r=-6h'-12h^2\,,\\
 r''+3hr'+\omega^2r=0\,.
 \end{cases}
\end{equation}
Introducing the new function $q_1\equiv r'$, we obtain the system of three first-order linear
differential equations:
\begin{equation}\label{sys_rdot_r_h}
\begin{cases}
 q_1'=-\cfrac{3}{2\tau}\,q_1-\omega^2 r\\
r' = q_1\\
h_1'=-\cfrac{1}{6}\,r-\cfrac{2}{t}\,h_1
\end{cases}
\end{equation}
Its characteristic polynomial is
\begin{equation}\label{polyn_2}\nonumber
 P(\tilde \lambda)=\left(\tilde \lambda+\frac{2}{\tau}\right)
 \left(\tilde\lambda^2+\frac{3\tilde \lambda}{2\tau }+\omega^2\right)\,,
\end{equation}
so the approximate eigenvalues for $\omega\tau\gg 1$ are
\begin{equation}\label{eigenval_2}
\tilde \lambda_1 = -\frac{2}{\tau}\,, \qquad\qquad \tilde\lambda_{2,3} \approx -\frac{3}{4\tau}\pm i\omega\,.
\end{equation}
As above, the solutions of the system (\ref{sys_rdot_r_h}) are linear combinations of eigenvectors $\tilde V_j$:
\be
[h_1, r, q_1] = \sum \tilde C_j \tilde V_j \exp \left[ \int^\tau d\tau'\,\tilde\lambda_j (\tau') \right]\,,
\label{h1-z1-y1_2}
\ee
where
\be \nonumber
\tilde V_1 = [1,\, 0,\, 0]\,, \qquad
\tilde V_{2,3} = \left[-\frac{1}{6(5/(4\tau )\pm i\omega )},\, 1,\, -\frac{3}{4\tau} \pm  i\omega \right]
\ee
and
\be
\exp \left[ \int^\tau d\tau'\,\tilde\lambda_1 (\tau') \right] \sim 1/\tau^2\,\,\,\, {\rm and}\,\,\,
\exp \left[ \int^\tau d\tau'\,\tilde\lambda_{2,3} (\tau') \right] \sim  e^{\pm i\omega \tau} /\tau^{3/4}\,.
\label{exp-lambda_2}
\ee
So the oscillating solution for $h_1$ is the same as that of eq. (\ref{eq:analytical_solution}). However, the slowly
varying (non-oscillating) solution for $h_1$ decreases more slowly, namely 
as $1/\tau^2$ instead of $1/\tau^3$.

Probably this difference is related to the freedom in the zeroth order GR solution, 
with respect to the shift of time:
\be
h_0 = \frac{1}{2(\tau +\delta)} =  \frac{1}{2\tau} - \frac{\delta}{2\tau^2 }\,.
\label{shift-h0}
\ee
Such freedom tells us that the terms of order $1/\tau^2$ in the first-order corrections are in some sense arbitrary.
So the solution $h_1 \sim 1/\tau^2$ is spurious and should be disregarded. Anyhow the
non-oscillating solutions for $h_1$ quickly disappear asymptotically and can be neglected.

The  solutions found above describe the oscillations of the Hubble parameter around the GR value $1/(2\tau)$. Moreover,
the amplitude of such oscillations decreases more slowly than $1/\tau$, so at some stage the oscillations
will become large and the condition $h_1\ll h$ will no longer be satisfied. After this stage is reached, the
linear approximation is no longer valid and the method used above becomes inapplicable.


However, we can still find the asymptotics of the exact nonlinear equations (\ref{R-of-H}) and (\ref{trace}) at large times
looking for solutions in the form:
\begin{subequations}\label{expansion_H_R}
\begin{equation}
h(\tau) = A(\tau) + B_s(\tau) \sin \nu\tau + B_c (\tau) \cos \nu\tau\,,
\label{H-of-t}
\end{equation}
\begin{equation}
r(\tau) = C(\tau) + D_s (\tau) \sin \nu \tau + D_c (\tau) \cos \nu \tau\,.
\label{R-of-t}
\end{equation}
\end{subequations}
Here coefficients $A,B,C$, and $D$ are slowly varying functions of time and $\nu$, assumed large, may in principle be different from $\omega$. 
As we will see below, this is indeed the case due to radiative corrections (scalaron mass renormalization).
In the approximation taken here we find $\nu=\omega $. 

We obtain approximate equations for these functions equating the coefficients in front of the slowly varying terms, and in front of 
 $\sin \omega \tau$ and $\cos \omega \tau$. These equations are approximate because we do not take into
account higher-frequency terms which appear as a result of non-linearity, but the taken approximation happens to be quite accurate.
Doing so, we find from equations (\ref{trace_dimensionless}):
\begin{subequations}\label{eqn-ABCD}
\begin{align}
&A' + 2 A^2 +B_s^2 + B_c^2 = -C/6\,,\label{eqn-ABCD_A}\\
&B_s' - B_c \nu + 4 A B_s = -D_s/6\,,\label{eqn-ABCD_Bs}\\
& B_c' + B_s \nu + 4 A B_c = -D_c/6\,,\label{eqn-ABCD_Bc}\\
&C'' + 3\left(A C' + \frac{1}{2} B_s  D_s'  + \frac{1}{2} B_c D_c'
 -  \frac{1}{2}\nu  B_s  D_c  +  \frac{1}{2}\nu  B_c  D_s\right) + \omega^2 C = 0\,,\label{eqn-ABCD_C}\\
&D_s'' - 2 \nu D_c' -\nu^2 D_s  + 3A (D_s' - \nu D_c )
 +3 C' B_s + \omega^2 D_s = 0\,, \label{eqn-ABCD_Ds}\\
&D_c'' + 2 \nu D_s' -\nu^2 D_c  + 3A (D_c' + \nu D_s )
 +3 C' B_c + \omega^2 D_c = 0\,.\label{eqn-ABCD_Dc}
\end{align}
\end{subequations}
Assuming
\begin{equation}\label{expand_ABCD}
A = \frac{a}{\tau},\,\, B_s = \frac{b_s}{\tau},\,\,\  B_c = \frac{b_c}{\tau},\,\,\,
C=  \frac{c}{\tau^2},\,\, D_s = \frac{d_s}{\tau},\,\,\,  D_c = \frac{d_c}{\tau}\,,
\end{equation}
and keeping only the dominant terms (lowest powers of $1/\tau $) we obtain the solutions
\begin{subequations}\label{sol_ABCD}
\begin{align}
  &\nu^2=\omega^2\,,\label{sol_ABCD_omega}\\
 &d_s=6\omega b_c\,,\label{sol_ABCD_Ds}\\
&d_c=-6\omega b_s\,,\label{sol_ABCD_Dc}\\
&b_s^2+b_c^2=2a(2a-1)\,,\label{sol_ABCD_B}\\
&c=18a(1-2a)\label{sol_ABCD_C}\,.
 \end{align}
\end{subequations}
It is interesting that equation (\ref{sol_ABCD_B}) demands $a>1/2$ in presence of oscillations.
We will see from the numerical solution that this is indeed the case.

To summarize, the situation is the following: we initially assume $h_1/h\ll 1$ and solve the
linearized systems (\ref{sys:linear_perturb}) or (\ref{sys_rdot_r_h}), finding that the Hubble parameter
oscillates around the value $1/(2\tau)$ with amplitude growing with time as $h_1/h\sim \tau^{1/4}$; eventually,
such oscillations (and hence non-linear terms) would become dominant and the linear approximation
becomes invalid. However, we can proceed further using a sort of truncated Fourier expansion which
allows to take into account the non-linearity of the system in the limit
$\omega\tau\gg 1$. As a result we have found that $h_1/h\rar$ const. In other words, the amplitude of the
oscillating part of $h$ asymptotically behaves as $1/\tau$, i.e. in the same way as the slowly-varying part of $h$.


To be completely sure about these analytical results we have to check if the exact numerical solution of the system (\ref{sys:hubble_evolution})
shows the same behavior. Still, the analytical estimates presented above are of great interest for the
calculation of the evolution of $R$ and $H$ when particle production effects are taken into account.

\subsubsection{Numerical Solutions \label{ss-num-estim}}

We integrate the system of equations (\ref{sys:hubble_evolution}) starting at $\tau_0=1/2$, with the initial conditions
\be\label{eq:initial_conditions_2}
 \begin{aligned}
  &h_0=1+\delta h_0\\
  &h'_0=-2+\delta h'_0\\
 &y_0=1+\delta y_0\,,
 \end{aligned}
\ee
where $\delta h_0$, $\delta h'_0$ and $\delta y_0$ do not vanish simultaneously. As expected, numerical integration with the initial conditions given by
eq.~(\ref{eq:initial_conditions_1}) gives the usual GR solution $h=1/(2\tau)$ within numerical precision, so we are interested in the more general case in
which the initial conditions deviate from the GR values.

As it was said earlier, systems (\ref{sys:hubble_evolution}) and (\ref{sys:curvature}) are equivalent. 
However, for the numerical integration of these systems one has to specify initial values of different 
quantities. For the integration of system (\ref{sys:hubble_evolution}) one has to fix  $h_0$, $h'_0$, and $y_0$,
while for the integration of (\ref{sys:curvature}) the values of $h_0$, $r_0$, and $r'_0$ must be specified.
The expression of one set of initial values through the equivalent values of another set can be found
using the equations under scrutiny.   
Indeed, once $h_0$, $h'_0$ and $y_0$ are chosen, $h''_0$ is uniquely determined
through the first equation in (\ref{sys:hubble_evolution}), and consequently $r_0$ and $r'_0$ are specified as well, via (\ref{sys:curvature}). After all, both
systems are equivalent to the same single third-order differential equation, whose solution is determined by three initial conditions.

\begin{figure}[!h]
\centering
\includegraphics[width=.4\textwidth]{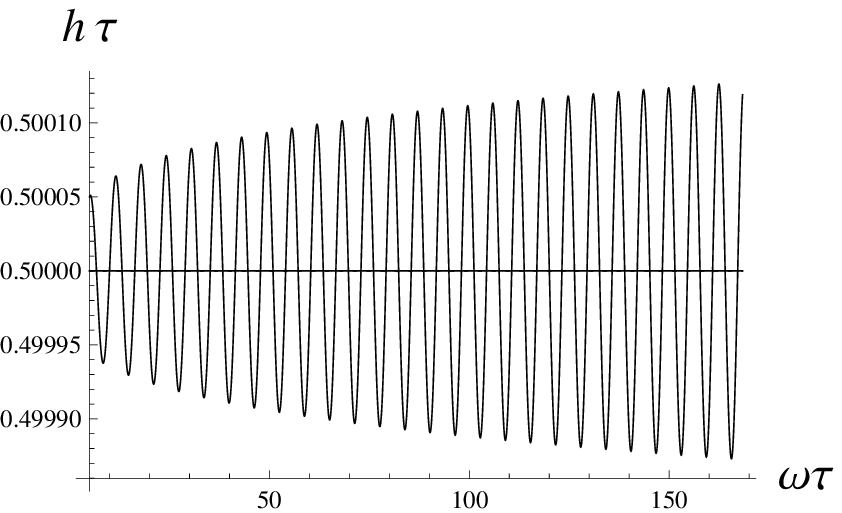}
\includegraphics[width=.4\textwidth]{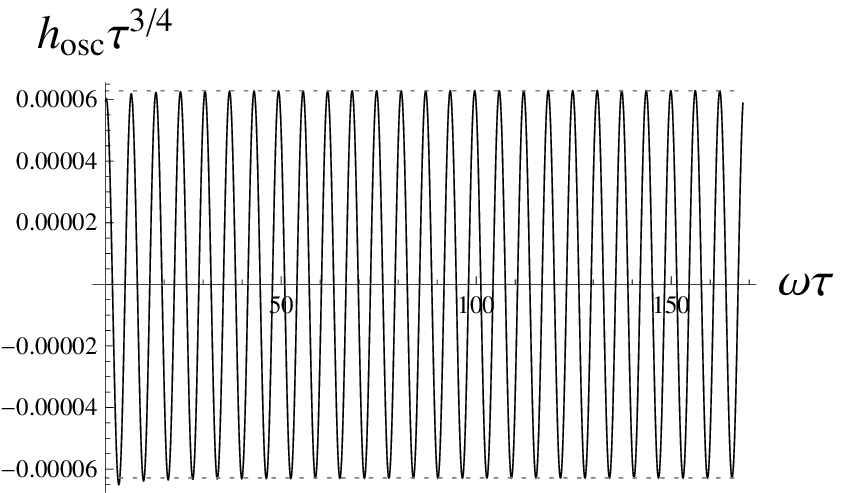}
\caption{Numerical solution of eqs.~(\ref{sys:hubble_evolution}) with $\delta h_0=10^{-4}$, $\delta h'_0=0$, $y_0=1$,
and $\omega=10$.
The best fit, for functional form (\ref{h_fit_func}), is given by $\alpha\simeq 1$, $\beta\simeq 6.29\times 10^{-5}$.}
\label{fig:h0=1.0001}
\end{figure}

\begin{figure}[!h]
\centering
\includegraphics[width=.4\textwidth]{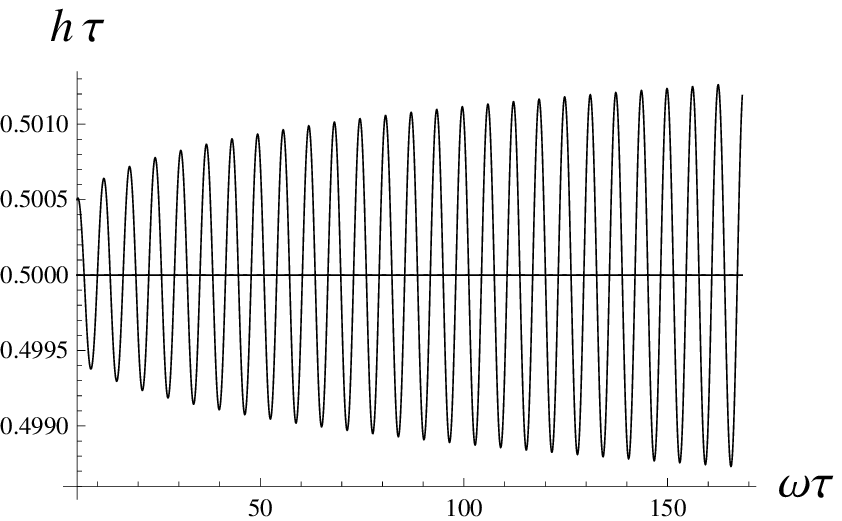}
\includegraphics[width=.4\textwidth]{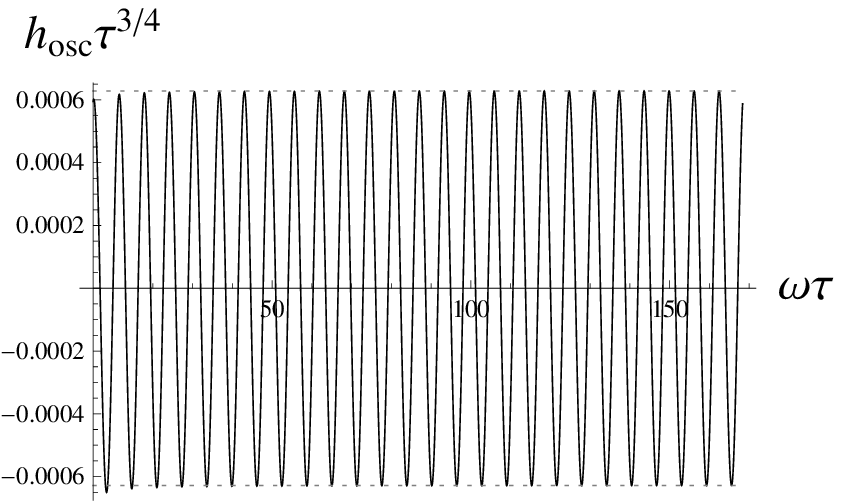}
\caption{Numerical solution of system (\ref{sys:hubble_evolution}) with $\delta h_0=10^{-3}$, $\delta h'_0=0$, $y_0=1$, $\omega=10$. The best fit is $\alpha\simeq 1$, $\beta\simeq 6.28\times10^{-4}$.}
\label{fig:h0=1.001}
\end{figure}

\begin{figure}[!h]
\centering
\includegraphics[width=.4\textwidth]{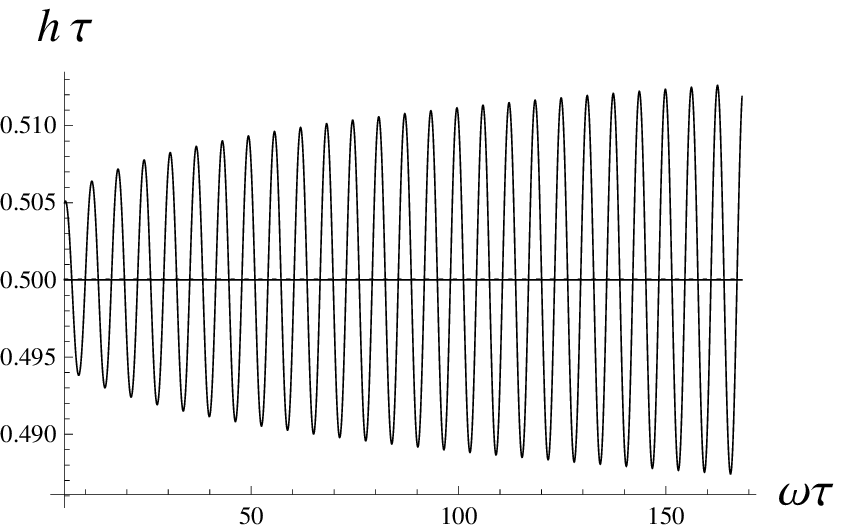}
\includegraphics[width=.4\textwidth]{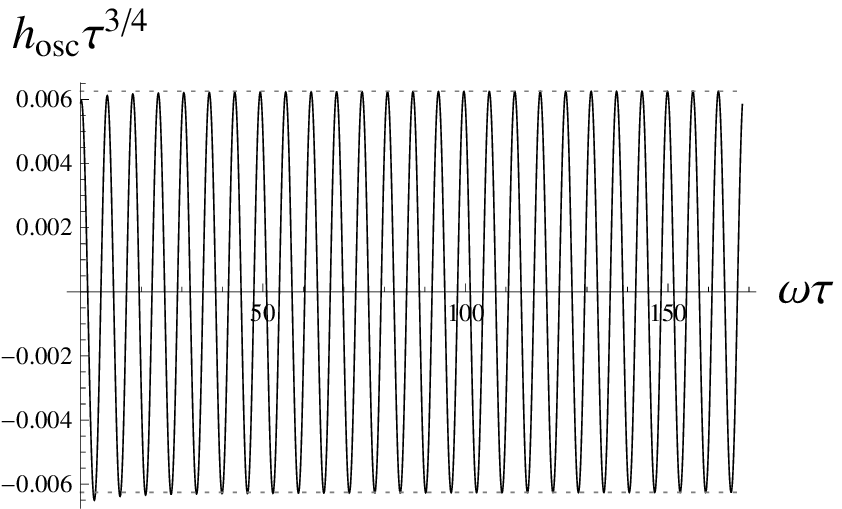}
\caption{Numerical solution of  system (\ref{sys:hubble_evolution}) with $\delta h_0=10^{-2}$, $\delta h'_0=0$, $y_0=1$, $\omega=10$. The best fit is given by $\alpha\simeq 1$, $\beta\simeq 6.26\times 10^{-3}$.}
\label{fig:h0=1.01}
\end{figure}

\begin{figure}[!h]
\centering
\includegraphics[width=.4\textwidth]{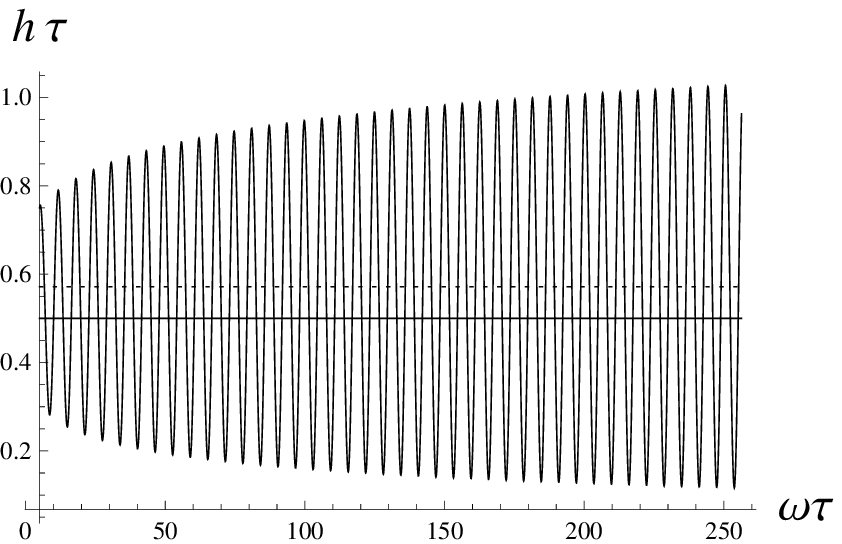}
\includegraphics[width=.4\textwidth]{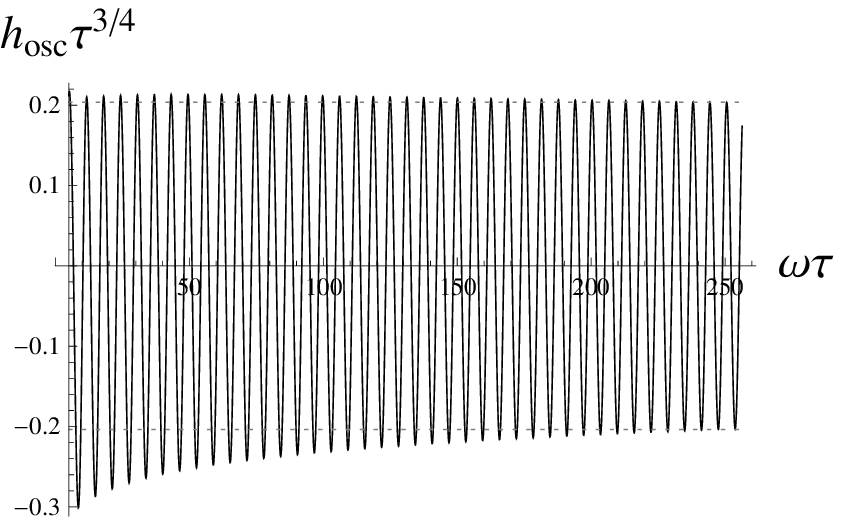}
\caption{Numerical solution of system (\ref{sys:hubble_evolution}) with $\delta h_0=0.5$, $\delta h'_0=0$, $y_0=1$, $\omega=10$. The best fit at large times is $\alpha\simeq 1.14$ (dotted line in the left panel), $\beta\simeq 0.20$. Please also note, in the right panel, that the oscillating part of $h$ does not decrease \textit{exactly} as $\tau^{-3/4}$. The apparent up-down asymmetry in $h_{osc}$ is due to the fact that $\alpha$ is a function of time, and that we centered oscillations around its value at late times.}
\label{fig:h0=1.5}
\end{figure}

\begin{figure}[!h]
\centering
\includegraphics[width=.4\textwidth]{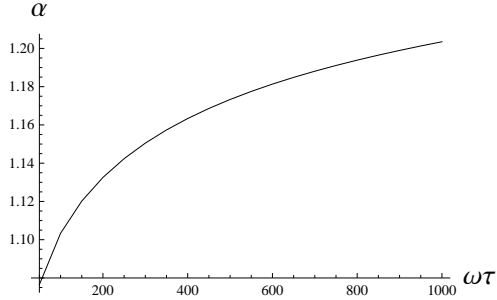}
\caption{Evolution of $\alpha$ with time. Initial conditions are those of figure \ref{fig:h0=1.5}.}
\label{fig:alpha_1.5}
\end{figure}

We have found that the numerical solutions of the system of equations~(\ref{sys:hubble_evolution}, \ref{eq:initial_conditions_2}) are in
very good agreement with the previous analytical estimates in the linear regime, i.e. for initial conditions fulfilling the
requirements  $\delta h_0/h_0\ll 1$, $\delta y_0/y_0\ll 1$, and $\omega\tau\gg 1$. In figures \ref{fig:h0=1.0001}, \ref{fig:h0=1.001},
and \ref{fig:h0=1.01} we present the numerical results for the dimensionless Hubble parameter $h$ determined
from the system of equations~(\ref{sys:hubble_evolution}), with $\omega=10$ and initial conditions
\begin{subequations}
\begin{align}
 \delta h_0 = 10^{-4}&\,, \quad \delta h'_0=0\,,\quad \delta y_0=0\qquad \text{(fig. \ref{fig:h0=1.0001})}\\
\delta h_0 = 10^{-3}&\,,\quad \delta h'_0=0\,,\quad \delta y_0=0\qquad \text{(fig. \ref{fig:h0=1.001})}\\
\delta h_0 = 10^{-2}&\,,\quad \delta h'_0=0\,,\quad \delta y_0=0\qquad \text{(fig. \ref{fig:h0=1.01})}
\end{align}
\end{subequations}
The function $h\tau$ is found to oscillate around the central value
$h=1/2$ with amplitude $h_1\sim\tau^{-3/4}$. As the deviation from the ideal GR behaviour increases, 
the average value of $h\tau$ also varies, and in general it
is no longer equal to 1/2. A very good functional form for fitting the solutions is
\begin{equation}\label{h_fit_func}
h(\tau)\simeq \frac{\alpha + \beta\,\tau^{1/4}\sin(\omega\tau+\varphi)}{2\tau}\equiv \frac{\alpha}{2\tau}+h_{osc}\,,
\end{equation}
where the dimensionless parameters $\alpha$ and $\beta$ are very slowly varying functions of time. This fit is shown, for instance,
in figures \ref{fig:h0=1.5} and \ref{fig:alpha_1.5}, where the numerical solution for $\delta h_0=0.5$ is presented.
A deviation from the analytical estimates of the linearized equations is to be expected in this situation, since the condition $\delta h_0/h_0 \ll 1$ 
is not fulfilled and non-linear terms in eqs.~(\ref{sys:hubble_evolution}) are important.

For the moment, let us concentrate on the case of small $\delta h_0$, in which we can safely take $\alpha=1$. Qualitatively,
one notices that parameter $\beta$, evaluated at the same value of $\omega\tau$, increases roughly linearly with the initial displacement $\delta h_0$.
Moreover, when solving the system of equations (\ref{sys:hubble_evolution}) with $\delta h_0=0$ and varying $\delta h'_0$, we find again a
(roughly) linear relation of the form $\beta\propto\delta h'_0$.
\vspace{-.5cm}

\begin{figure}[!h]
\centering
\includegraphics[width=.4\textwidth]{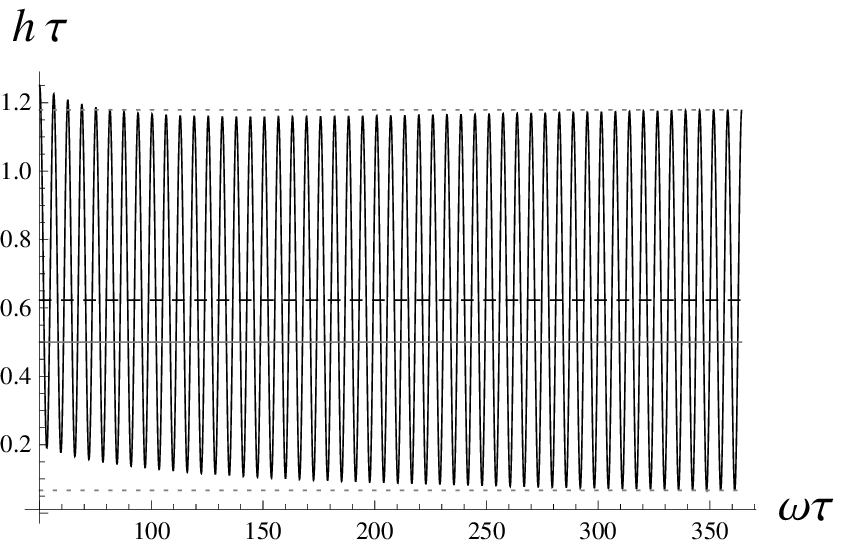}
\includegraphics[width=.4\textwidth]{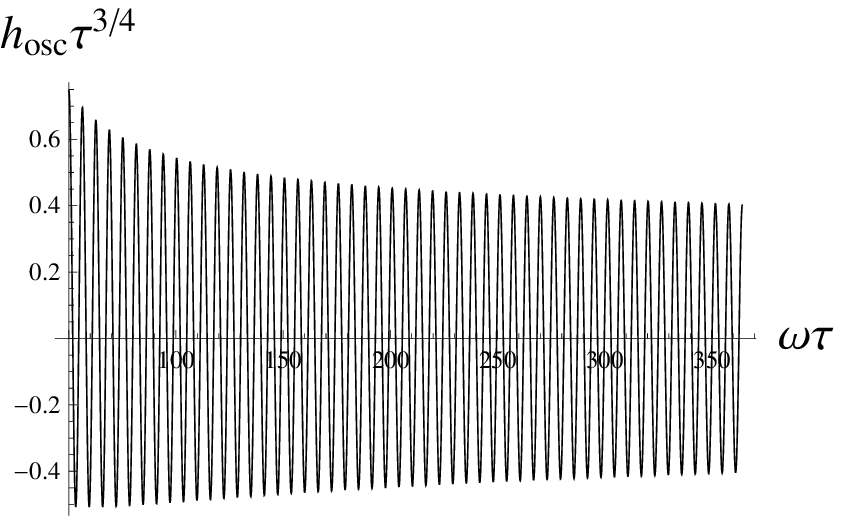}
\caption{Numerical solution for the dimensionless Hubble parameter $h$ with initial conditions (\ref{initial_cond_2.5}).
Oscillations in the left panel are almost of constant amplitude, whereas in the right panel they are clearly decreasing.}
\label{fig:h0_2.5}
\end{figure}

\begin{figure}[!h]
\centering
\includegraphics[width=.4\textwidth]{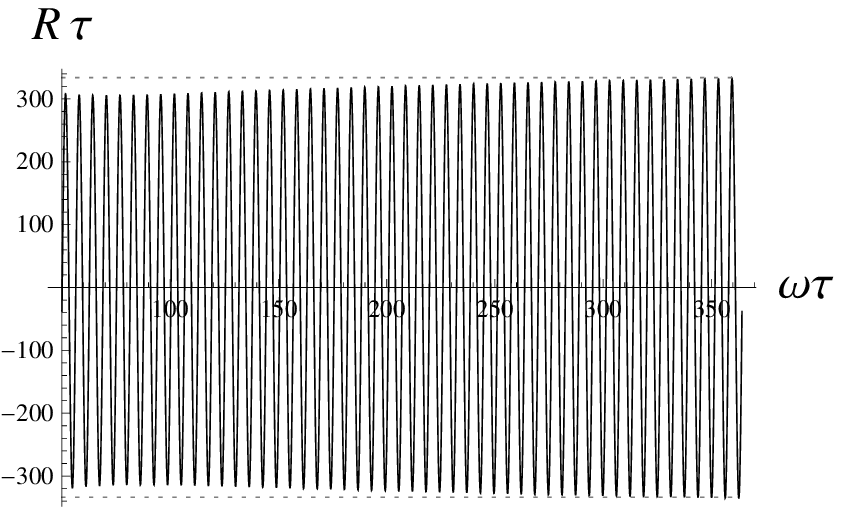}
\includegraphics[width=.4\textwidth]{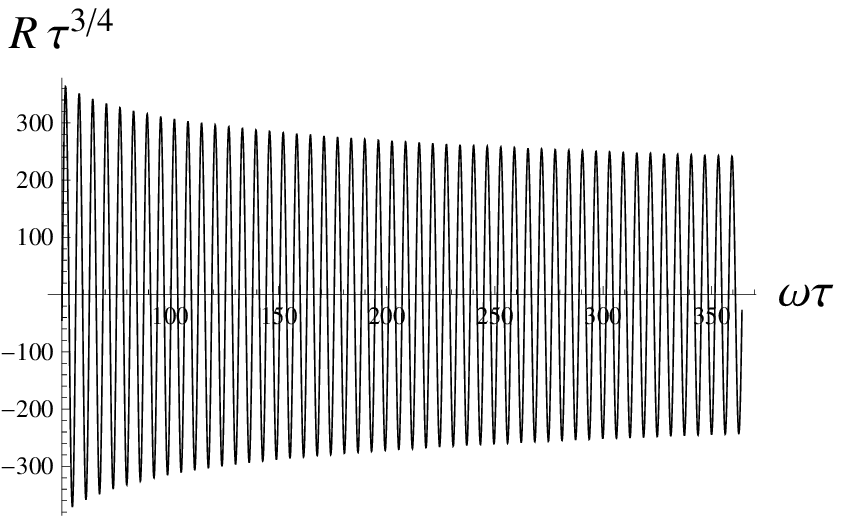}
\caption{Numerical solution for the dimensionless scalar curvature $r$ with initial conditions (\ref{initial_cond_2.5}).}
\label{fig:r0_2.5}
\end{figure}

In figures \ref{fig:h0_2.5} and \ref{fig:r0_2.5} we present the numerical results for the initial conditions
\begin{equation}\label{initial_cond_2.5}
  \delta h_0=1.5\,,\quad \delta h'_0=0\,,\quad\delta y_0=0\,,\quad\omega=100\,.
\end{equation}
Results are, at least qualitatively, in agreement with the analytical estimates made in the non-linear regime, for $\omega\tau\gg 1$. Evidently, the amplitudes of the oscillating terms of both $h$ and $r$ decrease faster than $\tau^{-3/4}$ (linear regime), and rather close to $\tau^{-1}$. Furthermore, the Hubble parameter does not oscillate around the GR value $h\tau=1/2$, but around a larger value, as expected from eq.~(\ref{sol_ABCD_B}).
In fact, for these initial conditions, the best fit at the considered final integration time is given by $ \alpha\simeq 1.246$.
The universe evolution in $R + R^2$ gravity after inflation, without relativistic matter, was considered in ref.~\cite{Gur-Star}, where it was stated that the non-oscillating part of $h$ tends asymptotically to the GR matter-domination value $h\tau=2/3$, in contrast to our relativistic case. We plan to study what happens in cosmology with non-relativistic matter in the future.\\
Our result that $h_{osc}\sim t^{-3/4}$ (\ref{sol_h_linear}) in the linear regime agrees with what found in the last paper of \cite{Gur-Star}, namely that $h_{osc}\sim a^{-3/2}$. The latter, however, is not in perfect agreement with our results in the non-linear regime, where we have $\alpha\simeq 1.246$ but $h_{osc}\sim 1/\tau$. This is probably due to the fact that we have not been able to reach the true asymptotic regime, but only a pre-asymptotic region.

\section{Particle Production and Back-Reaction} \label{s-part-prod}

The particle production rate by the oscillating gravitational field in $R^2$ gravity was considered in ref.~\cite{Starobinsky_1980,Vilenkin_1985},
where it was estimated as $\Gamma \sim m^3/m_{Pl}^2$.
In this section we present more rigorous calculations, which are essentially in agreement with ref.~\cite{Vilenkin_1985}.
We will derive below a closed equation of motion for the cosmological evolution of
$R$ with the account of the back-reaction of particle production. To this end we consider
a massless scalar field $\phi$ minimally-coupled to gravity. Its action can be written as:
\begin{equation}\label{phi_action}
 S_\phi=\frac{1}{2}\int d^4x\,\sqrt{-g}\,g^{\mu\nu}\partial_\mu\phi\,\partial_\nu\phi\,.
\end{equation}
In spatially-flat FRW background (\ref{FRW}) it leads to the equation of motion:
\begin{equation}\label{eq_motion_phi}
 \ddot \phi+3H\dot\phi-\frac{1}{a^2}\Delta\phi=0\,.
\end{equation}
Field $\phi$ enters the equation of motion for $R$ (\ref{trace}) via the trace of its energy-momentum tensor:
\[
T^\mu_\mu (\phi)=-g^{\mu\nu}\partial_\mu\phi\,\partial_\nu\phi\equiv -(\partial\phi)^2\,.
\]
It is convenient to introduce the conformally rescaled field, $\chi\equiv a(t)\phi$, and the conformal time $\eta$,
such that $a\,d\eta=dt$. It terms of these quantities we can rewrite the equations of motion as:
\be
 R''+2\cfrac{a'}{a}R'+m^2a^2R=8\pi\cfrac{m^2}{\mpl^2}\cfrac{1}{a^2}\left[\chi'^2-(\vec\nabla\chi)^2+\cfrac{a'^2}{a^2}\chi^2-\cfrac{a'}{a}(\chi\chi'+\chi'\chi)\right]\,,
 \label{R-diprime}
 \ee
\be
R=-6a''/a^3\,,
\label{R}
\ee
\be
 \chi''-\Delta\chi+\cfrac{1}{6}\,a^2R\,\chi=0\,,
\label{chi-diprime}
\ee
while action (\ref{phi_action}) takes the form:
\begin{equation}\label{chi_action}
 S_\chi = \frac{1}{2}\int d\eta\,d^3x\,\left(\chi'^2-(\vec\nabla\chi)^2-\frac{a^2R}{6}\chi^2\right)\,.
\end{equation}
Here and above prime denotes derivative with respect to conformal time.

Our aim is to derive a closed equation for $R$ taking the average value of the $\chi$-dependent quantum operators
in the r.h.s. of eq.~(\ref{R-diprime}) over vacuum, in presence of an external classical gravitational field $R$.
Our arguments essentially repeat those of ref.~\cite{Dolgov_Hansen}, where the
equation was derived in one-loop approximation.

We quantize the free field $\chi^{(0)}$ as usual:
\begin{equation}\label{quantiz_chi}
 \chi^{(0)}(x)=\int\frac{d^3k}{(2\pi)^3\,2E_k}\left[\hat a_k\,e^{-ik\cdot x}+\hat a^\dagger_k\,e^{ik\cdot x}\right]\,,
\end{equation}
where $x^\mu=(\eta,\mathbf x)$, $k^\mu=(E_k,\mathbf k)$, and $k_\mu k^\mu =0$. The creation-annihilation
operators satisfy the usual Bose commutation relations:
\begin{equation}\label{commutator}
 \left[\hat a_k,\hat a_k^\dagger\right]=(2\pi)^3\,2E_k\,\delta^{(3)}(\mathbf k-\mathbf k').
\end{equation}
Equation  (\ref{chi-diprime}) has the formal solution
\begin{equation}\label{formal_sol}
 \chi(x)=\chi^{(0)}(x)-\frac{1}{6}\int d^4y\,G(x,y)\,a^2(y)R(y)\chi(y)\equiv \chi^{(0)}(x)+\delta\chi(x)\,,
\end{equation}
where the massless Green's function is
\begin{equation}\label{green_func}
 G(x,y)=\frac{1}{4\pi|\mathbf x-\mathbf y|}\delta\left((x_0-y_0)-|\mathbf x-\mathbf y|\right)\equiv \frac{1}{4\pi r}\delta(\Delta\eta-r)\,.
\end{equation}
We assume that the particle production effects slightly perturb the free solution, so that $\delta\chi$ can be considered small and the
Dyson-like series can be truncated at first order, yielding
\begin{equation}\label{approx_chi}
 \chi(x)\simeq \chi^{(0)}(x)-\frac{1}{6}\int d^4y\,G(x,y)\,a^2(y)R(y)\chi^{(0)}(y)\equiv\chi^{(0)}(x)+\chi^{(1)}(x)\,.
\end{equation}
We can now calculate the vacuum expectation values
of the various terms in the right-hand side of equation (\ref{R-diprime}),
keeping only first-order terms in $\chi^{(1)}$. All terms containing only $\chi^{(0)}$ and its derivatives have nothing to do
with particle production and can be re-absorbed by a renormalization procedure into the parameters of the theory,
so they are of little interest here. The other terms are calculated using formulas such as
\begin{subequations}
\begin{gather}
 \partial_x\int d^4y\,G(x,y)\,a^2(y)R(y)\chi(y)=\int d^4y\,G(x,y)\left[a^2R\partial_y\chi+\partial_y(a^2R)\chi\right]\,,\\
 \int_0^\infty dk\,e^{i\alpha k}=\pi\delta(\alpha)+i\mc P\left(\frac{1}{\alpha}\right)\,.
\end{gather}
\end{subequations}
We have collected  all the terms, arriving to the expressions:
\begin{subequations}\label{trace_energy_momentum_tensor_chi}
 \begin{align}
  &\langle\chi^2\rangle\simeq -\frac{1}{48\pi^2}\int^\eta_{\eta_0}d\eta'\,\frac{a^2(\eta')R(\eta')}{\eta-\eta'}\,,\label{chi_squared}\\
 &\langle \chi'^2-(\vec\nabla\chi)^2\rangle\simeq-\frac{1}{96\pi^2}\int^\eta_{\eta_0}d\eta'\,\frac{(a^2(\eta')R(\eta'))''}{\eta-\eta'}\,,\label{chi_prime_squared}\\
&\langle\chi\chi'+\chi'\chi\rangle\simeq -\frac{1}{48\pi^2}\int^\eta_{\eta_0}d\eta'\,\frac{(a^2(\eta')R(\eta'))'}{\eta-\eta'}\,.\label{chi_chi_prime}
 \end{align}
\end{subequations}
Substituting these expressions into (\ref{R-diprime}), we obtain a closed integro-differential equation for $R$, 
of which we will find an approximate analytical solution.
We also plan to find the numerical solution of this equation but this is a much more complicated problem. Still for our purpose the approximate
analytical solution is accurate enough.

First of all, one has to note that despite having oscillating $H$ and $R$, the scale factor $a$ basically follows a power-law expansion, so it varies very little during many oscillation times $\omega^{-1}$. Thus, we expect that $d\eta/\eta\sim dt/t$ and that the dominant part in the integrals in (\ref{approx_chi}) is given by derivatives of $R$, since $R'\sim \omega R+t^{-1}R\simeq\omega R$, because $\omega t\gg 1$.\\
The dominant contribution of particle production is therefore given by eq.~(\ref{chi_prime_squared}), and yields
\begin{equation}\label{R_with_back_reaction_approx}
 \ddot R+3H\dot R+m^2R\simeq -\frac{1}{12\pi}\frac{m^2}{\mpl^2}\frac{1}{a^4}\int_{\eta_0}^\eta d\eta'\,\frac{(a^2(\eta')R(\eta'))''}{\eta-\eta'}\simeq-\frac{1}{12\pi}\frac{m^2}{\mpl^2}\int_{t_0}^tdt'\,\frac{\ddot R(t')}{t-t'}\,.
\end{equation}
The equation is naturally non-local in time since the impact of particle production depends upon all 
the history of the evolution of the system. The equation is linear in $R$, in contradiction with reference~\cite{Mijic_Morris_Suen_1986}, 
where the r.h.s. of the equation is quadratic in $R$. The latter result is physically doubtful because if the 
sign of $R$ changes, the effect of the particle production term would not be a damping of oscillations but their amplification.

\subsection{Analytical Calculations}

We  repeat the calculations of section \ref{sec:analytical_estim_no_back_react} using  expansion (\ref{expansion_H_R}) and including
the back-reaction effects in the form of equation (\ref{R_with_back_reaction_approx}). The right-hand side of this equation can be
written as:
\begin{align}\label{eq:expansion_ddot_R_part_prod}
 g\int^t_{t_0}dt'\,\frac{\ddot R(t')}{t-t'}=\,&g\int_0^{t-t_0}d\tau\,\frac{\ddot R(t-\tau)}{\tau} = g\int_\epsilon^{t-t_0} d\tau\,\frac{\ddot R(t-\tau)}{\tau} + \int_0^\epsilon d\tau\,\frac{\ddot R(t-\tau)}{\tau}\notag \\
 =\,&g \int_\epsilon^{t-t_0}d\tau\,\frac{\ddot C}{\tau}+\notag\\
&+g \cos(m_1 t)\int_\epsilon^{t-t_0}d\tau\,\frac{1}{\tau}\left[F_c\cos(m_1\tau)-F_s\sin(m_1\tau)\right]+\notag\\
&+g\sin(m_1 t)\int_\epsilon^{t-t_0}d\tau\,\frac{1}{\tau}\left[F_c\sin(m_1\tau)+F_s\cos(m_1\tau)\right]\,,
\end{align}
where
\begin{subequations}
\begin{align}
&g\equiv -\frac{1}{12\pi}\frac{m_1^2}{\mpl^2}\,,\\
 & F_c \equiv \ddot D_c+2 m_1\dot D_s-m_1^2D_c\,,\\
&F_s \equiv \ddot D_s-2m_1\dot D_c-m_1^2D_s\,.
\end{align}
\end{subequations}
To avoid confusion we need to mention that $\tau$ here is simply the integration variable and is
different from the dimensionless time $\tau$ of the previous sections.

We have  introduced here the new notation $m_1$ which is equal to $m$ plus radiative corrections
specified below, and which corresponds to $\nu$ in eqs.~(\ref{H-of-t}, \ref{R-of-t}). The difference between
$m$ and $m_1$ is not essential under the integral but it should be taken into account in the l.h.s. of eq. (37), where
we should take $m_1$ instead of $m$.

Please note that the slowly-varying functions $C$, $D_s$ and $D_c$ inside the integrals are to be 
evaluated at $(t-\tau)$, and a dot denotes derivative with respect to (physical) time $t$, not $\tau$. Because of the $1/\tau$ factor,
the integral is logarithmically divergent, but this divergence can be
absorbed into the renormalization of mass, $m$, and coupling, $g$. So we separate the integral into two parts: one where $\tau$ goes from 0 to some small parameter $\epsilon$, which determines the normalization point at which
 the physical mass and coupling are fixed, and another, taken 
from $\epsilon$ to $(t-t_0)$, which gives corrections to
the physical qualities due to interactions. More details
can be found in ref.~\cite{Dolgov_Hansen}.

Equating the coefficients multiplying the slow varying terms, $\sin m_1 t$, and $\cos m_1 t$ 
in the same way as it has been done in sec.~\ref{sec:analytical_estim_no_back_react}, see eqs.~(\ref{eqn-ABCD}), we obtain 
the same first three equations~(\ref{eqn-ABCD_A}, \ref{eqn-ABCD_Bs}, \ref{eqn-ABCD_Bc}), where the effects of particle production do not directly appear, 
and the remaining three ones with the additional terms coming from eq.~(\ref{R_with_back_reaction_approx}), see also expansion (\ref{eq:expansion_ddot_R_part_prod}). 
The latter equations become integro-differential but they can be reduced to differential equations in the case of fast oscillations.
So the complete set of equations with the account of particle production has the form (for convenience we also include
the unchanged first three equations of set (\ref{eqn-ABCD}):
\begin{subequations}\label{eqn-ABCD_with_back_react}
\begin{align}
&\dot A + 2 A^2 +B_s^2 + B_c^2 = -C/6\,,\\
&\dot B_s - B_c m_1 + 4 A B_s = -D_s/6\,,\\
& \dot B_c + B_s m_1 + 4 A B_c = -D_c/6\,,\\
&\ddot C + 3A \dot C + \frac{3}{2} B_s \dot D_s + \frac{3}{2} B_c \dot D_c
 - \frac{3}{2}m_1 B_s D_c + \frac{3}{2}m_1 B_c  D_s + m^2 C \simeq g\int_\epsilon^{t-t_0}d\tau\,\frac{\ddot C}{\tau}\,,\label{eqn-ABCD_with_back_react_C}\\
&\ddot D_s +(m^2 - m_1^2) D_s
- 2 m_1 \dot D_c\,  +3A (\dot D_s - m _1 D_c) +\notag\\
& \qquad\qquad\qquad\qquad+3\dot C B_s \simeq g\int_\epsilon^{t-t_0}d\tau\,\frac{F_s\cos(m\tau)+F_c\sin(m\tau)}{\tau}\,,\label{eqn-ABCD_with_back_react_Ds}\\
&\ddot D_c +(m^2 - m_1^2) D_c
+ 2 m_1 \dot D_s\, + 3A (\dot D_c + m_1 D_s) +\notag\\
& \qquad\qquad\qquad\qquad+3\dot C B_c \simeq g\int_\epsilon^{t-t_0}d\tau\,\frac{F_c\cos(m\tau)-F_s\sin (m\tau)}{\tau}\,.\label{eqn-ABCD_with_back_react_Dc}
\end{align}
\end{subequations}
In integrals (\ref{eqn-ABCD_with_back_react_Ds}) and (\ref{eqn-ABCD_with_back_react_Dc}), containing quickly oscillating functions, the effective value of $\tau$ is about $1/m$. Thus we can approximate
$F(t-\tau) \approx F(t)$ and take such factors out of the integrals. Let us analyze now for example equation (\ref{eqn-ABCD_with_back_react_Ds}) term by term. The analysis
of eq.~(\ref{eqn-ABCD_with_back_react_Dc}) is similar. In what follows we neglect $\ddot D$ in comparison with $m^2 D$.

The dominant term in eq.~(\ref{eqn-ABCD_with_back_react_Ds}), which is the coefficient multiplying $D_s$, determines the renormalization of $m$:
\be
m_1^2 = m^2 + g\,m^2 \int_\epsilon^{t-t_0} \frac{d\tau}{\tau}\,\cos m\tau  \,.
\label{renorm-mass}
\ee
The next subdominant term, which is the coefficient in front of $\dot D_c$, determines the decay rate of $D_c$:
\be\label{dot-Dc}
\dot D_c=\frac{g m}{2}D_c\int_\epsilon^{t-t_0}\frac{d\tau}{\tau}\sin m\tau\approx \frac{\pi g m}{4}D_c\,.
\ee
We skipped here the term $g \dot D_c $, which leads to higher order corrections to the production rate.
Thus the decay rate is
\be
\Gamma_R = -\frac{\pi g m}{4} = \frac{m^3}{48m_{Pl}^2}\,.
\label{Gamma-R}
\ee
Correspondingly the oscillating part of $R$ or $H$ behaves as
\be
\cos m_1 t \rar e^{-\Gamma_R t} \, \cos m_1 t\,.
\label{damping}
\ee
We will use this result in the next subsection in the calculation of the energy density influx of the produced particles
into the primeval plasma.

\subsection{Gravitational Particle Production}

From equation (\ref{chi-diprime}) follows that  the amplitude of gravitational production of 
two identical $\chi$
particles with momenta $p_1$ and $p_2$ in the first order in perturbation theory is given by
\begin{equation}\label{particle_production_general}
A(p_1,p_2)\simeq \int d\eta\,d^3x\,\frac{a^2R}{6}\left\langle p_1,p_2\left|\chi\chi \right|0\right\rangle\,,
\end{equation}
where the final two-particle state is defined by
\[
 |p,q\rangle = \frac{1}{\sqrt 2}\,\hat a^\dagger_p\,\hat a^\dagger_q |0\rangle\,.
\]
The factor $1/\sqrt{2}$ is simply the correct normalization of the two-particle state due to the Bose statistics. Using eq.~(\ref{quantiz_chi}), we find
\begin{align}\label{chi2_p1_p2_amplitude}
 \langle p_1,p_2|\chi\chi|0\rangle &= \frac{\sqrt 2}{8(2\pi)^6}\int\frac{d^3k\,d^3k'}{E_k\,E_{k'}}e^{i(E_k+E_{k'})\eta-i(\mathbf k+\mathbf k')\cdot \mathbf x}\langle 0|\,\hat a_{p_1}\,\hat a_{p_2}\,\hat a^\dagger_k\,\hat a^\dagger_{k'}\,|0\rangle\,,\notag\\
&=\sqrt 2\,e^{i(E_{p_1}+ E_{p_2})\eta-i(\mathbf p_1+\mathbf p_2)\cdot\mathbf x}\,.
\end{align}
Here $E_k^2 =\mathbf k^2$, and the function $a^2R$ has the form
\[
 a^2(\eta) R(\eta)=D(\eta)\sin(\tilde\omega\eta)\,,
\]
where $D(\eta)$ is a slowly-varying function of (conformal) time, $\tilde\omega $ is the frequency conjugated to conformal time. 
Under these approximations, the amplitude (\ref{particle_production_general}) becomes
\begin{align}\label{calc_part_prod_general}
 A(p_1,p_2)&=\frac{i}{6\sqrt 2}\int d\eta\,d^3x\,D(\eta)\left(e^{i\tilde \omega \eta}-e^{-i\tilde\omega\eta}\right)e^{i(E_{p_1}+ E_{p_2})\eta}\,
 e^{-i(\mathbf p_1+\mathbf p_2)\cdot \mathbf x}\,.
\end{align}
Taking $E_{p_i}\geq 0$ and neglecting at this stage the variation of $D$ with time, we obtain
\begin{equation}\label{part_prod_2}
 A(p_1,p_2)\simeq -\frac{i}{6\sqrt 2}\,D(\eta)(2\pi)^4\,\delta^{(3)}(\mathbf p_1+\mathbf p_2)\,\delta(E_{p_1}+E_{p_2}-\tilde\omega)\,.
\end{equation}
In order to find the particle production
rate per unit comoving volume and unit conformal time, we need to integrate the modulus squared of this amplitude over the phase space, namely: 
\begin{align}\label{trans_rate}
 n' &= \int\frac{d^2p_1\,d^3p_2}{(2\pi)^6\,4\,E_{p_1}E_{p_2}}\,\frac{|A(p_1,p_2)|^2}{V\,\Delta\eta}\notag \\
 &\simeq\frac{1}{72}\,D^2(\eta)\int\frac{d^3p_1\,d^3p_2}{(2\pi)^6\,4\,E_{p_1}E_{p_2}}(2\pi)^4\,\delta^{(3)}
 (\mathbf p_1+\mathbf p_2)\,\delta(E_{p_1}+E_{p_2}-\tilde\omega) \\
&\simeq \frac{D^2(\eta)}{576\pi}\,,
\end{align}
where $V$ and $\Delta\eta$ are the total volume and conformal time, which of course go to infinity,
$n$ is the number density of the produced particles, and a prime denotes derivative with respect to conformal time.
Since the energy of the produced particles is equal to $\tilde\omega/2$,
we find for the rate of gravitational energy transformation into elementary particles:
\begin{equation}\label{rho_prime_part_prod}
\rho'= \frac{n' \tilde \omega}{2}=\frac{D^2(\eta)\tilde\omega}{1152\pi}
\end{equation}
and so the rate of variation of the physical energy density of the produced $\chi$-particles is
\begin{equation}\label{rho_dot_part_prod}
 \dot\rho_{\chi}= \frac{m\langle R^2\rangle}{1152\pi}\,.
\end{equation}
Here, $\langle R^2\rangle$ is the square of the amplitude of the oscillations of $R$ and we substituted $\tilde\omega=a m$. 
To obtain the  the total rate of the gravitational energy transformation into elementary
particles we should multiply the above result by the number of the produced particle species, $N_{eff}$,
so the total rate of production of matter is $\dot \rho_{PP} = N_{eff} \dot \rho_\chi $.\\
Note that  result (\ref{rho_dot_part_prod}) coincides with that of ref.~\cite{Vilenkin_1985},
although performed in a slightly different cosmological regime.

Now we can calculate the evolution of the cosmological energy density of matter,
 which is determined by the equation:
 \be
 \dot\rho =-4H\rho + \dot\rho_{PP}\,.
\label{dot-rho-rel}
\ee
We assumed here that the produced matter is relativistic and so the first term in the r.h.s.
describes the usual cosmological red-shift, while the second term is the particle source from the
oscillations of $R$. Since $\rho$ is not oscillating but a smoothly varying function of time, its
red-shift is predominantly determined by the non-oscillating part of the Hubble parameter,
$H_{c} \simeq \alpha/2t$, see eq.~(\ref{h_fit_func}).

Parametrizing the oscillating part of the Hubble parameter as $H_{osc} \simeq {\beta \cos mt}/{t}$, 
we find for the oscillating part of curvature:
\be
R\simeq -\frac{6\beta m\sin mt}{t}\, e^{-\Gamma_R t}\,.
\label{R-osc}
\ee
Here we took into account the exponential damping of $R$, which was for brevity omitted in 
the expression for $H$ just above.\\
Correspondingly the energy density of matter  obeys the equation:
\be
\dot \rho = -\frac{2 \alpha}{t} \rho + \frac{\beta^2 m^3 N_{eff}}{32\pi t^2}\,e^{-2\Gamma_R t}\,.
\label{dot-rho-2}
\ee
This equation can be explicitly integrated as it is, but for a simple analytical estimate we will use the
instant decay approximation. Namely we neglect the exponential damping term, when $2\Gamma_R t <1$,
and take $\alpha =\alpha_1 = 1.25$, according to the numerical estimate of sec.~\ref{ss-num-estim}. For
$2\Gamma_R t  > 1$ we completely ignore the second (source) term in eq.~(\ref{dot-rho-2}) and
take $\alpha = \alpha_2 = 1$. This choice corresponds to the GR 
solution and we believe that it is realized when the
oscillations disappear, as follows from the analytical estimates presented above. Thus at short times,
$2 \Gamma_R t <1$, the energy density of matter would be:
\be
\rho = \rho_{in} \left(\frac{t_{in}}{t}\right)^{2\alpha_1} +
\frac{\beta^2 m^3 N_{eff}}{32 \pi (2\alpha_1-1) t} \left( 1 - \frac{t_{in}^{2\alpha_1 -1}}{t^{2\alpha_1 -1}} \right)\,.
\label{rho-short-t}
\ee
For large times, i.e. $2\Gamma_R t >1$, equation (\ref{dot-rho-2}) becomes homogeneous and its solution is
simply the relativistically red-shifted energy density with the initial value determined from eq.~(\ref{rho-short-t})
at $t= 1/(2\Gamma_R)$:
\be
\rho = \frac{ m^6 }{768 \pi\,m_{Pl}^2\, (2\Gamma_R t)^{2\alpha_2}}\,
\left[ \frac{\kappa}{8} \left(2 t_{in} \Gamma_R \right)^{2\alpha_1 - 2} +
\frac{\beta^2 N_{eff}}{2\alpha_1-1}\left(1-(2\Gamma_R t_{in})^{2\alpha_1-1}\right) \right] \,,
\label{rho-long-time}
\ee
where we parametrized the energy density of matter at the initial time $t_{in} $ as
\be
\rho_{in} =\frac{3 m_{Pl}^2 \kappa}{32 \pi t_{in}^2}\,.
\label{rho-in}
\ee
Parameter $\kappa$ is arbitrary, and depends upon the thermal history of the universe before $t_{in}$.
In particular, $\kappa = 0$ is possible and does not contradict our picture, since the equations of motion
have non-trivial oscillating solutions even if $\rho =0$.

The first term in equation (\ref{rho-long-time}) is the contribution of normal thermalized relativistic matter,
while the second also describes relativistic matter, but this matter might not be thermalized, at least during
some cosmological period. Depending upon parameters the relative magnitude of non-thermalized matter
might vary from negligibly small up to being the dominant one.

\section{Discussion and Implications}

The characteristic decay time of the oscillating curvature is
\be
\tau_R = \frac{1}{2\Gamma_R} =  \frac{24 m^2_{Pl}}{m^3} \simeq 2
\left(\frac{10^5\text{ GeV}}{m}\right)^3 \text{ seconds}\,.
\label{tau-R}
\ee
The contribution of the produced particles into the total cosmological energy density reaches its maximum value
at approximately this time. The ratio of the energy density of the newly produced energetic particles and that of those already
existing in the plasma, according to eq.~(\ref{rho-long-time}), is:
\be
\frac{\rho_{hi}}{\rho_{therm}} = \frac{8\beta^2 N_{eff}}{\kappa(2\alpha_1-1)}\,\frac{1-(2\Gamma_Rt_{in})^{2\alpha_1-1}}{(2\Gamma_R t_{in})^{2\alpha_1-2}}\,.
\label{ratio}
\ee
If we take $t_{in} \simeq 1/m$, then $t_{in} \Gamma_R \simeq m^2/m_{Pl}^2 \ll 1$ and the effects of non-thermalized
matter may be negligible. However, for sufficiently large $\beta$ and possibly small $\kappa$ the non-thermal
particles may play a significant role in the cosmological history.\\
The influx of energetic protons and antiprotons
could have an impact on BBN. Thus this would either allow to obtain some bounds on $m$ or even to improve the agreement between the
theoretical predictions for BBN and the measurements of primordial light nuclei abundances.

The oscillating curvature might also be a source of dark matter in the form of heavy supersymmetric (SUSY)
particles. Since the expected light SUSY particles have not yet been discovered at LHC, to some people supersymmetry somewhat
lost its attractiveness. The contribution of the stable lightest SUSY particle into the cosmological energy
is proportional to
\be
\Omega \sim{ m^2_{SUSY} }/m_{Pl}
\label{Omega-SUSY}
\ee
and for $m_{SUSY} $ in the range $100-1000$ GeV the cosmological fraction of these particles would be
of order unity. It is exactly what is necessary for dark matter. However, it excludes thermally produced LSP's if they are much
heavier. If LSP's came from the decay of $R$ and their mass is larger than the ``mass'' of $R$, i.e. $m$,
the LSP production could be sufficiently suppressed to make a reasonable contribution to dark matter.

These and other manifestations of the considered scenario will be discussed elsewhere.

\section*{Acknowledgements}

This work was supported by the Grant of the Government of Russian Federation, No. 11.G34.31.0047.


\end{document}